\documentstyle[epsf]{mn}

\newif\ifAMStwofonts

\ifoldfss
  \ifCUPmtlplainloaded \else
    \NewTextAlphabet{textbfit} {cmbxti10} {}
    \NewTextAlphabet{textbfss} {cmssbx10} {}
    \NewMathAlphabet{mathbfit} {cmbxti10} {} 
    \NewMathAlphabet{mathbfss} {cmssbx10} {} 
  \fi
  \ifAMStwofonts
    \ifCUPmtlplainloaded \else
      \NewSymbolFont{upmath} {eurm10}
      \NewSymbolFont{AMSa} {msam10}
      \NewMathSymbol{\upi}     {0}{upmath}{19}
      \NewMathSymbol{\umu}     {0}{upmath}{16}
      \NewMathSymbol{\upartial}{0}{upmath}{40}
      \NewMathSymbol{\leqslant}{3}{AMSa}{36}
      \NewMathSymbol{\geqslant}{3}{AMSa}{3E}

      \let\leq=\leqslant 
      \let\geq=\geqslant 
    \fi
  \fi
\fi 

\ifnfssone
  \newmathalphabet{\mathit}
  \addtoversion{normal}{\mathit}{cmr}{m}{it}
  \addtoversion{bold}{\mathit}{cmr}{bx}{it}
  \newmathalphabet{\mathbfit} 
  \addtoversion{normal}{\mathbfit}{cmr}{bx}{it}
  \addtoversion{bold}{\mathbfit}{cmr}{bx}{it}
  \newmathalphabet{\mathbfss} 
  \addtoversion{normal}{\mathbfss}{cmss}{bx}{n}
  \addtoversion{bold}{\mathbfss}{cmss}{bx}{n}
  \ifAMStwofonts
    \ifCUPmtlplainloaded \else
      \UseAMStwoboldmath
      \makeatletter
      \new@mathgroup\upmath@group
      \define@mathgroup\mv@normal\upmath@group{eur}{m}{n}
      \define@mathgroup\mv@bold\upmath@group{eur}{b}{n}
      \edef\UPM{\hexnumber\upmath@group}
      \new@mathgroup\amsa@group
      \define@mathgroup\mv@normal\amsa@group{msa}{m}{n}
      \define@mathgroup\mv@bold\amsa@group{msa}{m}{n}
      \edef\AMSa{\hexnumber\amsa@group}
      \makeatother
      \mathchardef\upi="0\UPM19
      \mathchardef\umu="0\UPM16
      \mathchardef\upartial="0\UPM40
      \mathchardef\leqslant="3\AMSa36
      \mathchardef\geqslant="3\AMSa3E

      \let\leq=\leqslant 
      \let\geq=\geqslant 
    \fi
  \fi
\fi 

\ifnfsstwo
  \DeclareMathAlphabet{\mathbfit}{OT1}{cmr}{bx}{it}
  \SetMathAlphabet\mathbfit{bold}{OT1}{cmr}{bx}{it}
  \DeclareMathAlphabet{\mathbfss}{OT1}{cmss}{bx}{n}
  \SetMathAlphabet\mathbfss{bold}{OT1}{cmss}{bx}{n}
  \ifAMStwofonts
    \ifCUPmtlplainloaded \else
      \DeclareSymbolFont{UPM}{U}{eur}{m}{n}
      \SetSymbolFont{UPM}{bold}{U}{eur}{b}{n}
      \DeclareSymbolFont{AMSa}{U}{msa}{m}{n}
      \DeclareMathSymbol{\upi}{0}{UPM}{"19}
      \DeclareMathSymbol{\umu}{0}{UPM}{"16}
      \DeclareMathSymbol{\upartial}{0}{UPM}{"40}
      \DeclareMathSymbol{\leqslant}{3}{AMSa}{"36}
      \DeclareMathSymbol{\geqslant}{3}{AMSa}{"3E}

      \let\leq=\leqslant 
      \let\geq=\geqslant 
    \fi
  \fi
\fi 

\ifCUPmtlplainloaded \else
  \ifAMStwofonts \else 
    \def\upi{\pi}
    \def\umu{\mu}
    \def\upartial{\partial}
  \fi
\fi

\title{Survival and disruption of subsystems during a cold collapse}
\author[T. Tsuchiya]
       {T. Tsuchiya \\
        Department of Astronomy, Kyoto University, Kyoto 606-8502, Japan;
	tsuchiya@kusastro.kyoto-u.ac.jp
}
\date{}

\pagerange{\pageref{firstpage}--\pageref{lastpage}}
\pubyear{1998}

\begin{document}

\maketitle

\label{firstpage}

\begin{abstract}
Cold collapse of a cluster composed of small identical clumps, each of
which is in virial equilibrium, is considered. Since the clumps have no
relative motion with respect to each other initially, the cluster
collapses by its gravity. At the first collapse of the cluster, most of
the clumps are destroyed, but some survive. In order to find the
condition for the clumps to survive, we made systematic study in
two-parameter space: the number of the clumps $N_c$ and the size of the
clump $r_v$. We obtained the condition, $N_c \gg 1$ and $n_k \geq 1$,
where $n_k$ is related to $r_v$ and the initial radius of the cluster
$R_{ini}$ through the relation $R_{ini}/r_v = 2 N_c^{(n_k+5)/6}$. A
simple analytic argument supports the numerical result.  This $n_k$
corresponds to the index of the power spectrum of the density
fluctuation in the cosmological hierarchical clustering, and thus our
result may suggest that in the systems smaller than $2/(\Omega h^2)$Mpc,
the first violent collapse is strong enough to sweep away all
substructures which exist before the collapse.
\end{abstract}


\section{Introduction}

Hierarchical structures are seen in various astrophysical objects in the
universe, such as globular clusters in a galaxy, or galaxies in a
cluster of galaxies. Such a subsystem in a virialised system forms a
distinct object which is in virial equilibrium by itself. How such
subsystems form and evolve in a larger scale environment is a fundamental
problem of astronomy.

This is because, in general, mutual gravitational interaction tends to
destroy the subsystems. Once a composite system of a large halo and many
subsystems forms, there are several destructive processes, such as tidal
heating caused by close encounters among subsystems or between a
subsystem and a core of the halo\cite{spit58}. For the case of globular
clusters in a galaxy, bulge shocking, disc shocking \cite{ostr72} and
stellar evaporation\cite{spit71} are also effective.

The rate of the disruption of subsystems depends on the properties of
the subsystems and the environment. In the Galaxy, for example, a
cluster should have a size and mass in a certain narrow range in order
to survive for a Hubble
time\cite{fall77,capu84,cherno87,okaz95}. Moreover, the survival factor
depends on the distance from the Galactic centre\cite{vesp97}. More than
a half of the present globular clusters in the Galaxy will be destroyed
in the next Hubble time \cite{agui88,gned97}.

For clusters of galaxies, the main effective processes to disrupt
galaxies are encounters between the galaxies and the tidal heating by
the encounter with the cluster core. Richstone \& Malumath
\shortcite{rich83} calculated the evolution of clusters of galaxies,
taking into account the tidal stripping and mergers of the galaxies with
a Monte-Carlo method and showed that most galaxies could survive until a
Hubble time, despite large mass loss. By using the numerically
determined cross section for a merger of two identical galaxies, Makino
\& Hut\shortcite{makino-j97} estimated the merger rate of galaxies in a
cluster, and concluded that a galaxy experiences mergers, on average,
once every 100 Gyr. On the other hand, some numerical simulations
\cite{barn89,funa93} show that mergers are very effective in a cluster
with a relatively small number of galaxies ($\la100$); consequently the
galaxies will disappear rapidly in a Hubble time.

It might be noted that these studies considered the evolution of the
virialised subsystems in a completely virialised cluster. This is not the
case, however, in a hierarchical clustering scenario, since smaller
systems form earlier so that galaxies form earlier than clusters of
galaxies. Hence the galaxies in a cluster experienced the early violent
phase of the cold collapse of the cluster. Also for globular clusters,
there is a theory that globular clusters forms during the collapse of a
protogalaxy \cite{fall85}. In both cases, substantial subsystems should
have destroyed in the dynamically evolving phase of the parent system.

Though destructive processes in subsystems should be more effective in a
collapsing system than in a virialised one, only a few studies have been
made for the former and we have not yet arrived at a fundamental
understanding. This is partly because analytic approaches are
inapplicable to dynamically evolving systems. In addition, this is a
difficult problem also for numerical calculations, since the existence
of compact subsystems requires high resolution in mass and time, which
means that we need a large number of particles and a small time step to
cope with such systems. As a result it takes huge computational time. A
pioneering study of this problem was made by van Albada
\shortcite{vana82}. He studied cold collapses numerically with a wide
variety of initial conditions in order to make clear the origin of the
observed light distribution of elliptical galaxies. In his simulations,
however, clumps are so large that their sizes are comparable to the mean
separation of the clumps. Hence all clumps would be disrupted by the
cold collapse. Later, the importance of the present issue was recognised
in cosmological simulations of structure formation. In the numerical
simulations, substructures are swept away in clusters of galaxies
\cite{white-s87,fren88}. It is called the ``overmerging
problem''. Moore, Katz \& Lake \shortcite{moor96a} showed that current
numerical resolution is sufficient to prevent dissolution by two-body
relaxation, and that physical effects, such as tidal heating, are
important. They also argued that the size and the density of the
innermost core of the subsystems are crucial factors for their survival,
and the current numerical resolution may not be enough to resolve the
overmerging problem.

In this paper we investigate the fundamental destruction mechanisms of
the subsystems, especially in the circumstance that the environment is
dynamically evolving, using both numerical and analytical approaches. We
employ a simple model, in which the system initially consists of
identical virialised clumps. The clumps have no relative motion, so that
the system collapses by gravity. This model is characterised by only two
parameters: the number of clumps in a cluster and the size of the clumps
relative to that of the cluster. We make a systematic numerical survey
in the two-parameter space.

In Section 2, we explain the initial conditions of our simulations. The
results of the simulations are presented in Section 3. A simple analytic
estimate of the condition for the clumps to survive is given in Section
4. Final section is devoted to the summary.

\section{Initial Conditions and Model Parameters}

Our main aim is to find the condition for the survival of substructures
during a cold collapse. Qualitatively, we expect that smaller clumps are
more likely to survive because their binding energies are large. In
addition, since the collisional effect between stars in a stellar system
becomes weakened as the number of the stars increases, we also expect
more clumps to survive as the number of clumps increases.

The dependence of the fraction of survival on the number of the initial
clumps is rather uncertain. In order to find a quantitative condition,
therefore we need to survey all the parameter space systematically by
adopting simple initial conditions.

The initial models are characterised only by two parameters; one is the
size of clumps and the other is their number. All clumps are identical
in mass and size, and all particles belong to one of the clumps.

Each clump is realised as an equilibrium Plummer model \cite{binn87};
the density distribution $\rho(r)$ is given by
\begin{equation}
\rho = \frac{3 m_1}{4 \pi r_1^3} \left[ 1 + {(\frac{r}{r_1})}^2
\right]^{-5/2},
\label{eq:Plummer} 
\end{equation}
where $r$ is the radius, $r_1$ is the core radius, and $m_1$ is the mass
of the clump. The potential and distribution function of a particle at
the position, {\bf x}, with the velocity, {\bf v}, are given,
respectively, by
\begin{equation}
\Phi (r) = - \frac{G m_1}{\sqrt{r_1^2 + r^2}},
\end{equation}
and
\begin{equation}
f(\mbox{\bf x, v}) =
  K |\varepsilon |^{7/2},
\end{equation}
where $\varepsilon$ is the specific energy of a particle,
$\varepsilon = \frac{1}{2}v^2+\Phi(r)$, $G$ is the gravitational
constant, and $K$ is a numerical constant.

Since the smallest structure and shortest time scale are determined by
those of the clump, we take the mass, size, and dynamical time of the
clump as the units of computation. Specifically, $G = m_1 = r_v = 1$,
where $r_v$ is the virial radius defined by energy of the clump,
\begin{equation}
E = - \frac{G m_1^2}{4 r_v}. 
\end{equation}
In these units, the half mass radius is $r_h = 0.769$, and the crossing
time at $r_h$ is 0.834.

A cluster is composed of $N_c$ clumps, which are distributed uniformly
within a sphere. The centre of mass of each clump is located at a point
which is randomly selected from the sphere and has no bulk velocity
initially. Since the size of the clumps is fixed, the initial radius of
the cluster $R_{ini}$ is scaled by
\begin{equation}
  \frac{R_{ini}}{r_v} = 2 N_c^{(n_k + 5)/6},
\label{eq:RtoN}
\end{equation}
where $n_k$ is a parameter.  This relation corresponds to the ratio
between the radii of virialised objects in the Einstein-de Sitter
universe with the initial power spectrum such as $P(k)\propto k^{n_k}$
[e.g., \S 26 of Peebles \shortcite{peeb80}]. The larger $n_k$ becomes
the smaller relative size of the clumps. In particular, when $n_k = -3$
the mean density of the cluster is about the same as that of a clump.

In our simulation, the number of particles in a clump is fixed to be
512. With this number, the clump is stationary at least for $T\la
50$. In the longest run two-body relaxation may be effective, but
structural change is very small. Further, Moore et
al. \shortcite{moor96a} assure that for $N\ga100$ numerical effects on
disruption of clumps are negligible compared with physical effects such
as tidal heating. For confirmation, we made test runs with a half
($N=256$) and a double ($N=1024$) particles and found the difference in
the number of surviving clumps is within a factor of 15\%.

We examine the cases with four different values of $n_k$ : $n_k = -3$,
$-2$, $-1$, $0$, and for each $n_k$, we assign eight different values of
$N_c$ : $N_c = 2$, 4, 8, 16, 32, 64, 128, 256, except for $n_k$ =
0. Thus we have 31 different combinations of $n_k$ and $N_c$.  Further
we prepare two different realisations of random distribution of clumps,
which are referred to as RUN1 and RUN2. In Table 1, we summarised the
various parameters adopted in the present simulations. The first and
second columns are $n_k$ and $N_c$, respectively.The third, the fourth, and
the fifth columns are filling factor of clumps in the whole system,
the initial radius and virial ratio of the system, respectively. These
three parameters are determined according to $n_k$ and $N_c$.

In our numerical code, gravity is calculated by direct summation of
contribution of all particles using GRAPE-3A \cite{okum93}. The
softening length, $\epsilon$, and the time step of integration, $\delta
t$, are set to be $\epsilon = 1/32$ and $\delta t = 1/64$, respectively. 
The total energy of the system is conserved within an accuracy of a few
hundredth of 1\% on average and at maximum 1\%.

\section{Numerical study for survival condition}

\begin{figure*}
\epsfxsize=17cm
\epsfbox{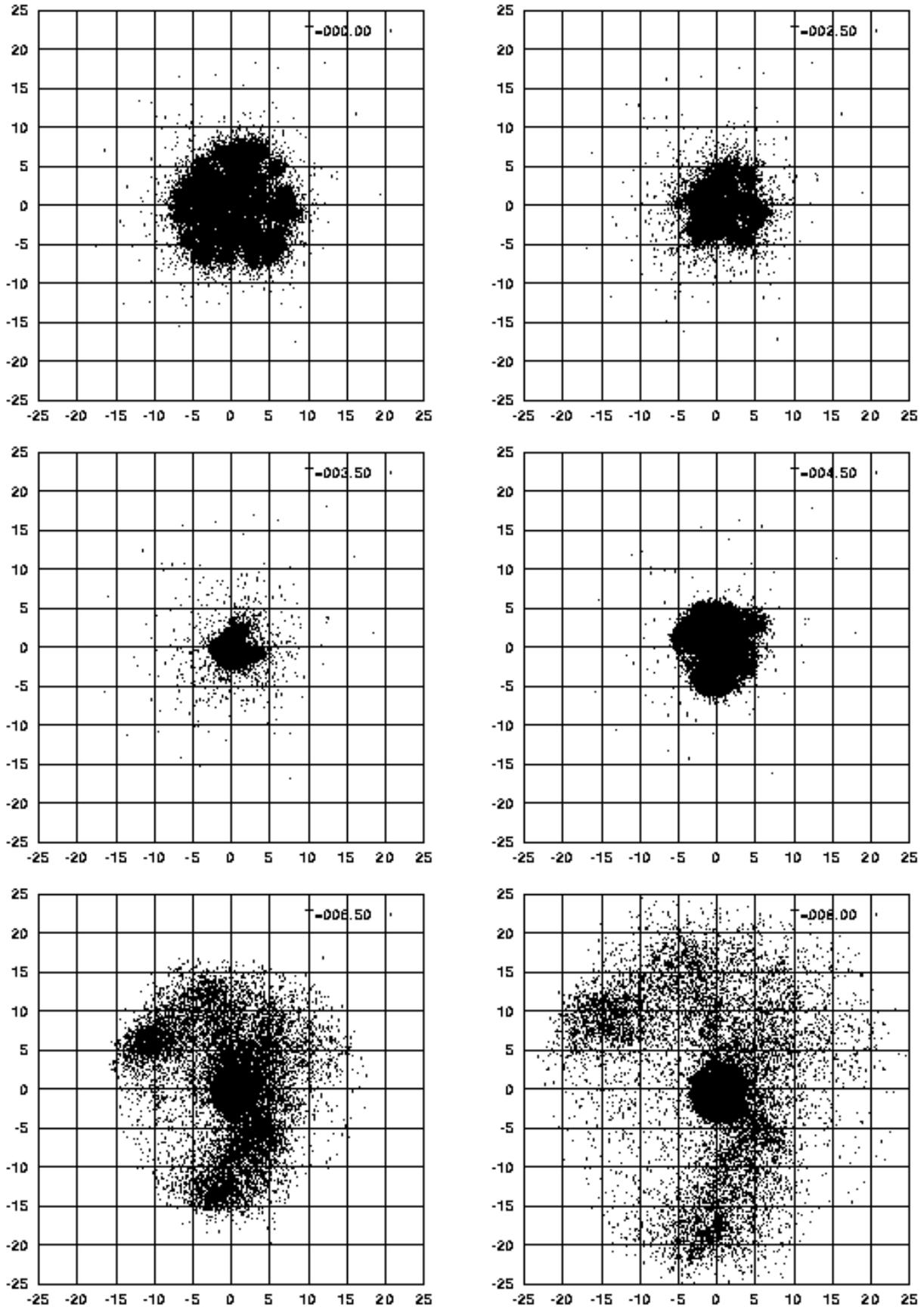}
\caption{Time evolution of the projected distribution in the $(y,z)$ plane
of the system with $n_k=-3$ and $N_c=64$ (example 1). The time is
indicated at the upper-right corner of each figure.}
\label{fig:snap1}
\end{figure*}

\begin{figure*}
\epsfxsize=17cm
\epsfbox{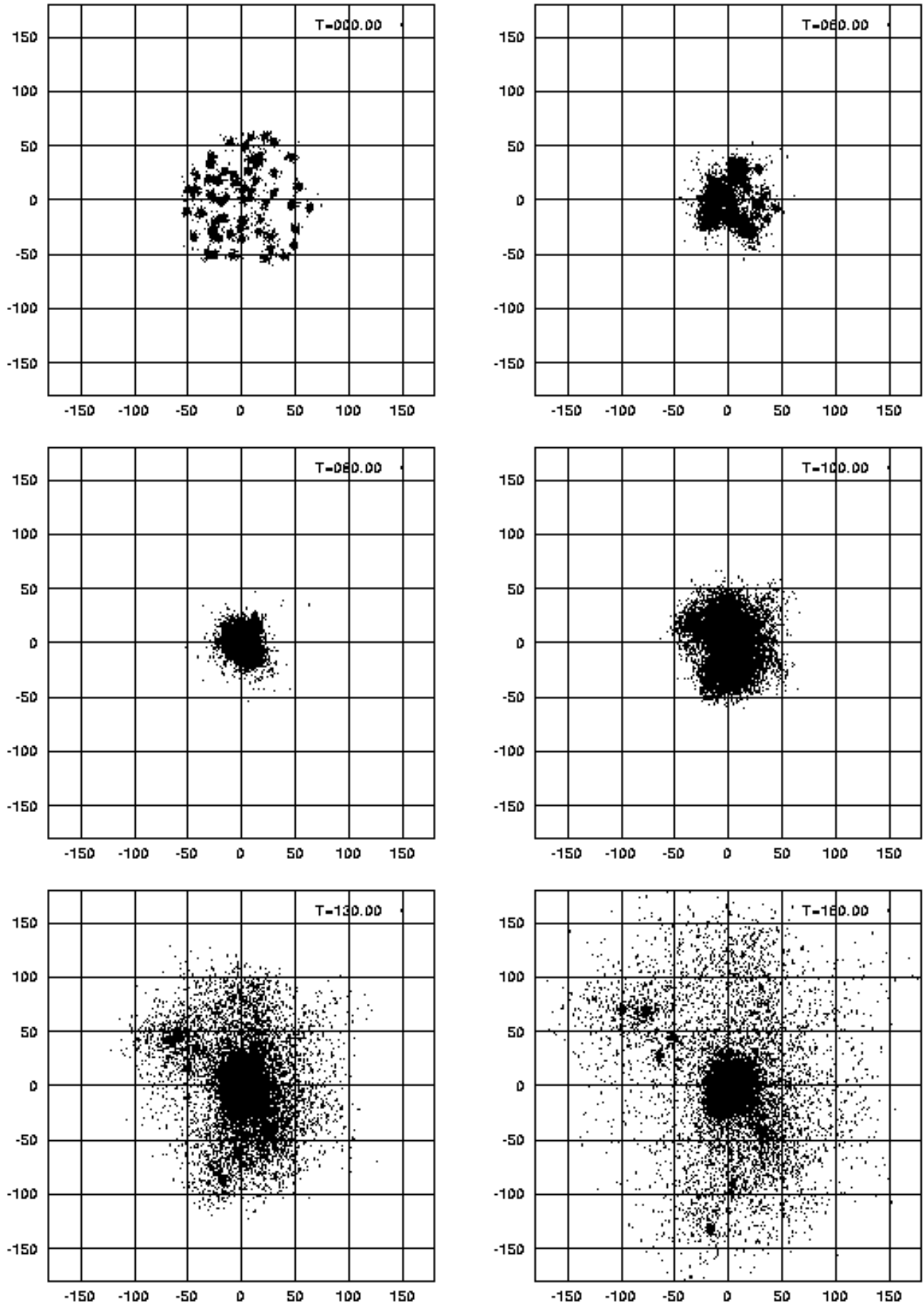}
\caption{Same as Fig.~\ref{fig:snap1} but the case with $n_k=0$ and
$N_c=64$ (example 2).}
\label{fig:snap2}
\end{figure*}

The cluster starts collapsing in a similar way to the case of a cold
collapse without substructure. Since the distribution of clumps in the
cluster is spherical and uniform, all the clumps collapse simultaneously
at $T = T_{coll}$. In this paper, we refer to this epoch as the first
collapse. The values of $T_{coll}$ for different parameters are listed
in Table~1. Figs.~\ref{fig:snap1} and \ref{fig:snap2} show two examples
of evolution : RUN1 with $n_k = -3$ and $N_c = 64$ (example 1), and RUN1
with $n_k = 0$ and $N_c = 64$ (example 2). We follow the evolution until
$T=2\,T_{coll}$, because we are concerned only with the first violent
stage. After $T=2\,T_{coll}$, the system no longer experiences violent
change in the potential, thus the processes of destruction of clumps are
the same as those which have already been studied by many authors
\cite{rich83,barn89,funa93,makino-j97}. The final state of our
simulation could be the initial state of the above studies. At the end
of our simulations, we count the number of the clumps which survive
through the first violent stage.

In the example 1, all clumps are destroyed at the first collapse and
merge into a large smooth halo. In the example 2, in contrast, many
clumps still have their identities even at $T=2\,T_{coll}$. (Note that
most of the survivors are in the central core, but they cannot be
resolved in Fig.~\ref{fig:snap2}.)

\begin{figure*}
\epsfxsize=14cm
\epsfbox{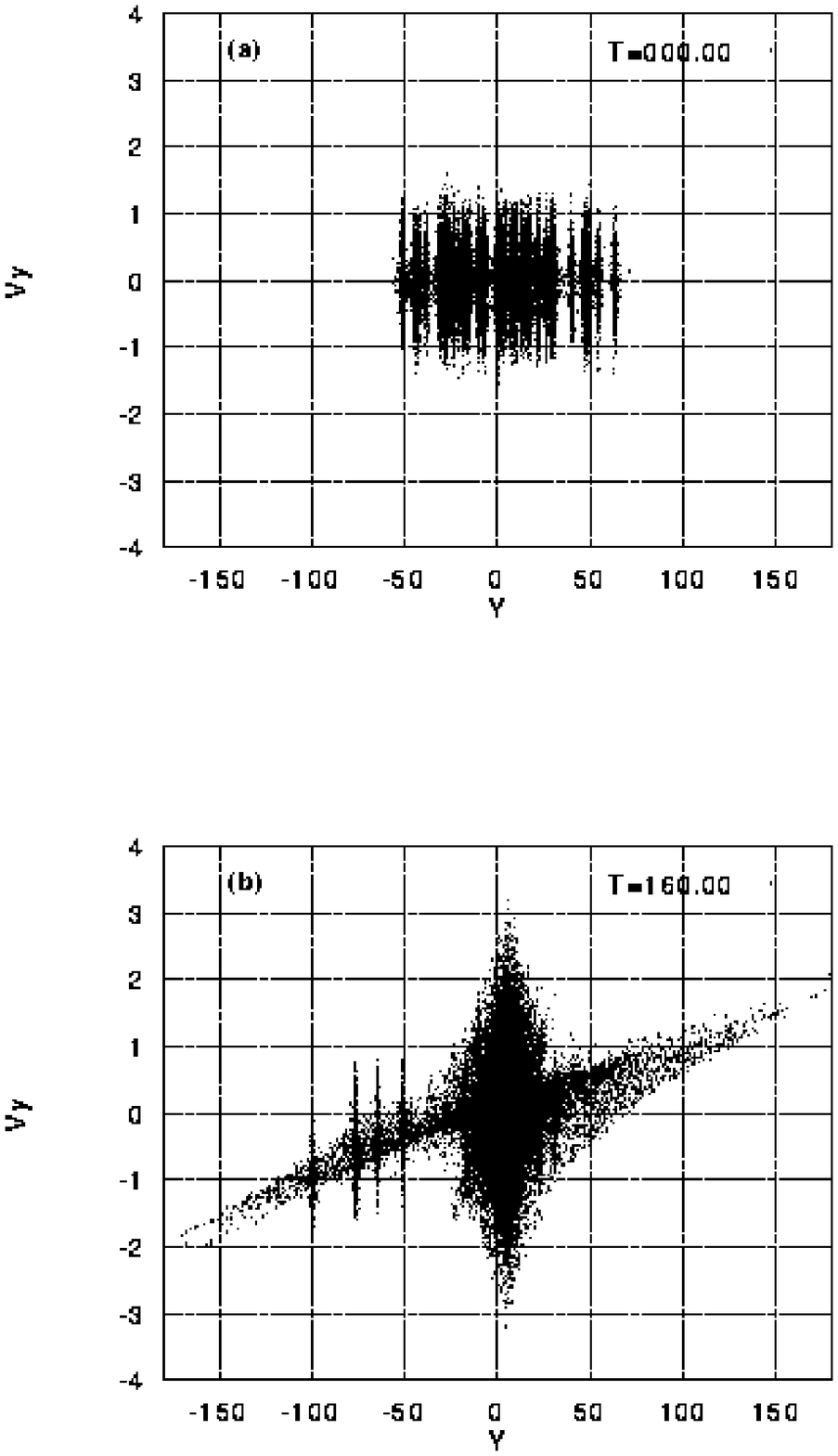}
\caption{Kinematical diagram of the systems with $n_k=0$ and $N_c=64$ in
the ($y$-$v_y$) plane (a)at the beginning ($T=0$), and (b) at the end of
the calculation ($T=160$).}
\label{fig:XVx}
\end{figure*}

Kinematical diagrams are much more useful to identify the surviving
clumps rather than density distribution. If a system is in a dynamical
equilibrium, the motion of a particle in the phase space is periodic so
that the trajectory should look like an oval or a spindle in a
position-velocity plane. Fig.~\ref{fig:XVx} shows the evolution of the
example 1, in the ($y$-$v_y$) plane, as a typical example. We examined
all three planes of the ($x$-$v_x$), ($y$-$v_y$), and ($z$-$v_z$) plane
and pick up the clumps with the spindle structure in the all three
plane.

\begin{figure*}
\epsfxsize=14cm
\epsfbox{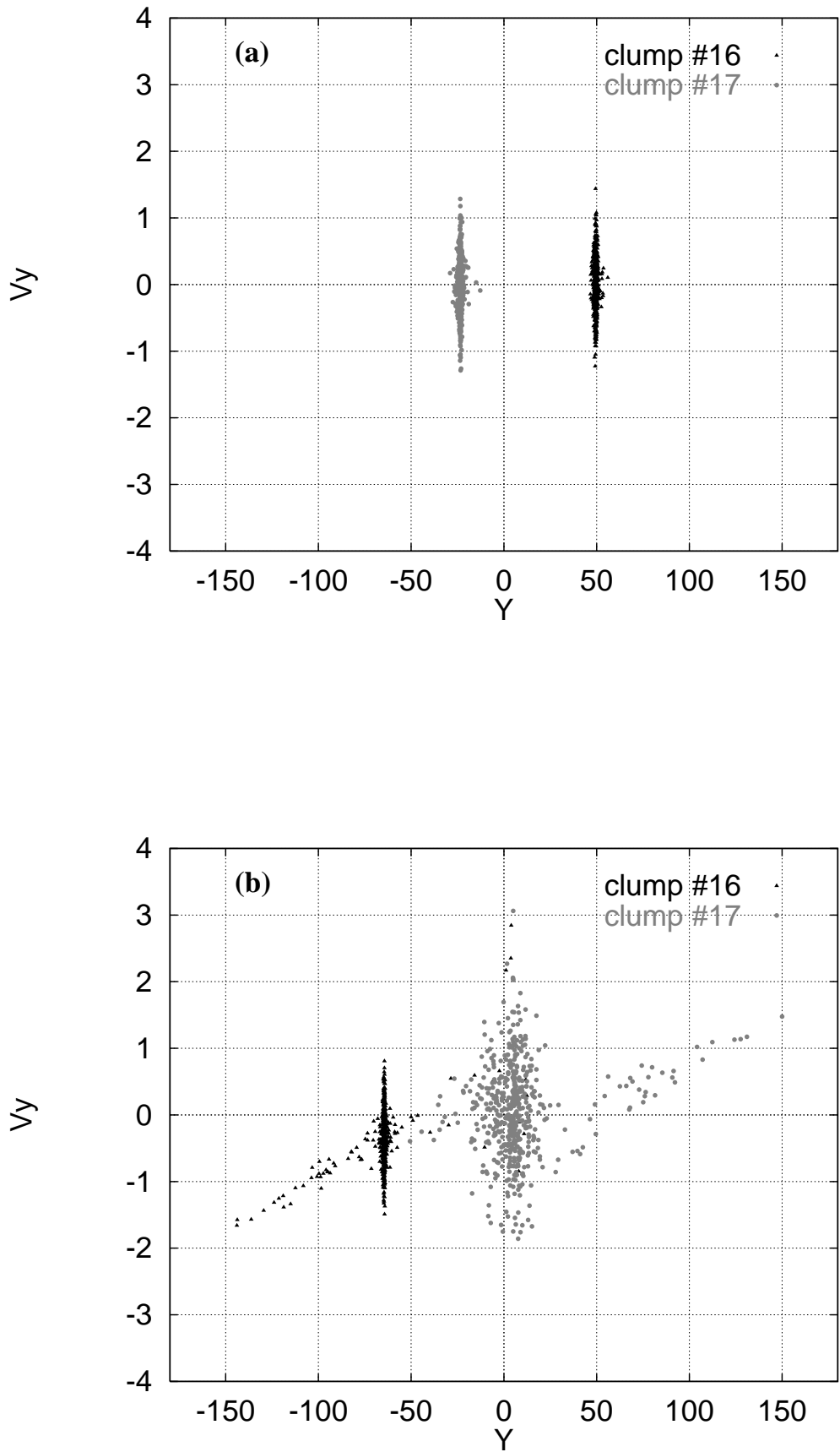}
\caption{Kinematical diagram of the same systems as in Fig. \ref{fig:XVx},
but showing only two clumps as an example. (a)At the beginning ($T=0$),
and (b) at the end of calculation ($T=160$). The structure of Clump \#16 
does not change much while Clump \#17 is destroyed and spreads over the
merged halo.}
\label{fig:clump}
\end{figure*}

At the beginning (see Fig.~\ref{fig:XVx}a) each clump forms an
individual spindle, whereas in the final state (see Fig.~\ref{fig:XVx}b)
many of the spindle structures are destroyed during the first collapse,
and a large merged halo is formed at the centre. There exist in the
latter two prominent features: the central big spindle structure, which
is due to the merged halo, and the rectilinear wings by escaping
particles from the merged halo. Besides, there are small spindle
structures superimposed on the main structures. These are due to the
surviving clumps. Fig. \ref{fig:clump} demonstrates the different
properties of the surviving and disrupted clumps. Two clumps are chosen
as representatives. Clump \#16 is destroyed in the first collapse and
spreads over the merged halo afterwards. Conversely, Clump \#17 still
retains the equilibrium structure a the end of calculation. We count the
number of all the surviving clumps after the calculation, and list them
in the column labelled $N_{\rm survive}$ of Table 1.

There are some clear tendencies in the results: first, irrespective of
the index $n_k$, the cluster with a small number of clumps ($N_c \leq
16$) yields no surviving clumps. Second, for $N_c \geq 32$ some clumps
may survive. The number of the survivors depends on $n_k$. For $n_k=-3$,
for example, all the clumps are destroyed except for the case with
$N_c=128$. For the larger $n_k$ and the larger $N_c$,the more clumps
survive.

One thing which we should note when considering the mechanism for the
disruption of the clumps is mass loss or escapers. It is known from the
studies of the cold collapses with uniform initial density distribution
without clumps, that models with a small initial virial ratio are liable
to suffer substantial mass loss. In the cases with clumpy initial
distribution, we observed even more mass loss. Among the escaping
particles, some clumps escape to the infinity and some not. It is thus
better to discriminate unbound survivors from bound ones in the central
big halo, since the subsequent evolution is distinct for those two.  As
unbound survivors, we pick up those which lie in the rectilinear wings
in the kinematical diagram. In Fig.~\ref{fig:XVx}, for example, four
such clumps can be seen in the wing extending to the negative $y$
direction. They are escaping from the central halo with a constant
velocity. Some other clumps are escaping in the $x$ and $z$ directions,
but still many survivors are found to orbit in the central halo. The
numbers of the bound survivors are listed in the column labelled $N_{\rm
bound}$ of Table 1. The number of the bound survivors increases as $n_k$
and $N_c$ increase. However, the largest difference between the
numbers of the bound and all survivors can be seen in $n_k=-2$. Only few
survivors are found to be bound, while there are many escaping
survivors. For $n_k=-1$ and 0, the number of bound survivors increases
as $N_c$ increases as long as $N_c\ga 64$.

Thus we now derive the conditions on the initial parameters for the
cluster to contain distinct clumps when it becomes virialised:
\begin{equation}
n_k \geq -1 \quad \mbox{and} \quad N_c \ga 64.
\end{equation}

\section{Simple analytic model}

In order to understand the physical mechanism underlying disruption of
clumps, we construct a simple analytic model.

Suppose that a cold cluster with $N_c$ clumps start to collapse by its
own gravity, same situation as those described in the previous
section. We surmise that the most effective mechanism for destroying
clumps is the tidal heating by the gravity of the whole cluster,
especially at the maximal collapse, where the radius of the system
becomes minimum. Let $m$ and $r_v$ be the mass and the the virial radius
of a clump, respectively, $M=N_cm$ be the total mass of the cluster, and
$R_{min}$ be the minimum radius of the cluster at the maximal
collapse. The clumps move in the gravitational potential of the cluster
with a typical distance $R_{min}$ and with a typical velocity $V \sim
(GM/R_{min})^{1/2}$. Then the change in the energy of a clump is given
by the tidal heating formula [e.g., eq.~(7-55) of Binney \& Tremaine
\shortcite{binn87}],
\begin{equation}
\Delta E_1 \approx \frac{G^2M^2m}{R_{min}^4V^2}r_v^2 \approx
\frac{GM}{R_{min}^3}mr_v^2 =
4\frac{M}{m}\left(\frac{r_v}{R_{min}}\right)^3 |E_1|,
\end{equation}
where $E_1$ is the energy of the clump given by
\begin{equation}
E_1=-\frac{Gm^2}{4r_v}.
\end{equation}

In the above formula, the minimum radius $R_{min}$ is an unknown
parameter. Here we employ the result of Aarseth, Lin \& Papaloizou
\shortcite{aars88}; in the cold collapse of a sphere with a uniform
distribution of $N$ particles, the system contracts to the minimum
radius $R_{min}$, depending on the number of particles,
\begin{equation}
R_{min}= N^{-1/3}R_{ini},
\label{eq:Aarseth}
\end{equation}
where $R_{ini}$ is the initial radius of the sphere. Since this minimum
radius results from the statistical $N^{1/2}$ fluctuation, we can apply the
result to our system by substituting the number of clumps $N_c$ for $N$
in equation (\ref{eq:Aarseth}). Moreover, using equation (\ref{eq:RtoN})
as a relation between $R_{ini}$ and $r_v$, the relative change in the
energy of the clump can be expressed only in terms of $N_c$ and $n_k$:
\begin{equation}
\frac{\Delta E_1}{|E_1|} = 4 N_c \left[ N_c^{-1/3} \cdot 2 N_c^{(n_k+5)/6} 
\right]^{-3} = \frac{1}{2} N_c^{-(n_k+1)/2}.
\label{eq:rele}
\end{equation}
If $n_k+1$ is positive, the relative change in the energy decreases as
$N_c$ increases. When $\Delta E_1/|E_1|$ is sufficiently small, a large
fraction of clumps are expected to survive the first collapse.

Surprisingly, this simple estimate is in excellent agreement with the
results of the numerical simulations, especially for the number of the
bound survivors. Therefore, we conclude that the disruption of the
clumps is mainly due to the tidal shock at the first collapse, and that
the conditions on the initial parameter for the cluster to sustain the
substructure until reaching the virial equilibrium are
\begin{equation}
n_k \geq -1 \quad \mbox{and} \quad N_c \gg 1.
\end{equation}

Note the estimate (\ref{eq:rele}) is valid for those clumps which
collide with each other simultaneously at the first collapse. Owing to
the statistical fluctuations, there exist some clumps which collapse
belatedly. Those clumps are accelerated to a larger velocity, and some
may gain enough energy to escape to the infinity. When the delayed
clumps reach their pericentre, however, the cluster as a whole has
already proceeded into the reexpansion phase, and thus the gravitational
force is weaker than that at the maximal collapse. Therefore, the
escaping clumps are more likely to survive. In fact, we found in the
simulations a tendency that the escaping survivors are initially located
in the outer part of the cluster. The bound survivors have a similar
tendency as well.

\section{Conclusions and discussion}

We have obtained the condition for substructures to survive after the
whole system has collapsed to reach a virial equilibrium. With the
number of clumps, $N_c$, and the power index $n_k$ of the spectrum of
initial density fluctuation as two parameters characterising the initial
distribution, the condition is expressed as
\begin{equation}
n_k \geq -1 \quad \mbox{and} \quad N_c \gg 1.
\end{equation}
As an application of this result, let us consider, for example, the CDM
model of cosmological clustering. It is known that in the CDM model the
power spectrum of the density fluctuation [eq. (25.22) of
Peebles\shortcite{peeb93}] is given by
\begin{equation}
P(k) \propto \frac{k}{\left[ 1+8k/\Omega h^2 + 4.7 k^2/(\Omega h^2)^2
\right]^2},
\end{equation}
and the gradient of the power spectrum is larger than $n_k=-1$ for the
fluctuation with the scale larger than $2/(\Omega h^2)$Mpc. Therefore
our result may suggest that in a system smaller than $2/(\Omega
h^2)$Mpc, the first violent collapse is strong enough to sweep away all
substructures which exist before the collapse.

Since our simple initial mass model has, in fact, no spectrum in the
distribution of masses of subsystems, it is more useful to express this
condition in terms of other parameters. Using radii of the cluster and
the clumps, for example, we derive
\begin{equation}
N_c \gg 1 \quad \mbox{and} \quad \frac{R_{ini}}{r_v} \ga 2N_c^{2/3}.
\end{equation}

For a system with many clumps, the filling factor $C$, which is the
inverse of the fraction of the volume occupied by the clumps, is widely
used to describe the clumpiness. In our case, $C= R_{ini}^3/(N_c
r_v^3)$, so that the condition is
\begin{equation}
N_c \gg 1 \quad \mbox{and} \quad C \ga 8N_c.
\end{equation}

Since our numerical model is quite idealised, we need to incorporate
many factors for more realistic discussion. First, as the model of each
clump, we examine only the Plummer sphere given by equation
(\ref{eq:Plummer}). This model has a core whose radius is about a half
of the half mass radius, $r_h$. On the other hand, the observed globular
clusters or elliptical galaxies have core radii of about $1/10$ or
$1/100$ of the half mass radii \cite{binn87}. Such smaller and denser
cores are expected to survive even when haloes or envelopes are
stripped. Second factor is rotation or asymmetry. Both make the first
collapse of the cluster less dense, and thus more clumps should survive
after the collapse of the cluster. All these factors weaken the
destructive effect. Therefore, the condition, presented in this paper,
for substructure to survive should be sufficient condition.

\section*{Acknowledgments}

I would like to thank Shunsuke Hozumi and Shin Mineshige for careful
reading of the manuscript, Junichiro Makino, Shogo Inagaki, and Eliani
Ardi for critical discussions. Financial support was provided by JSPS
Research Fellow.  The numerical simulations were run on a ``handmade''
GRAPE board at Department of Astronomy, Kyoto University, which was
assembled by Chiharu Ishizaka.



\begin{thebibliography}{}

\bibitem[\protect\citename{Aarseth, Lin \& Papaloizou }1988]{aars88}
Aarseth A. J., Lin D. N.~C., Papaloizou J. C.~B., 1988,
ApJ, 324, 288

\bibitem[\protect\citename{Aguilar, Hut \& Ostriker }1988]{agui88}
Aguilar L., Hut P., Ostriker J. P., 1988,
ApJ, 335, 720

\bibitem[\protect\citename{Barnes }1989]{barn89}
Barnes J.~E., 1989,
Nature, 338, 123

\bibitem[\protect\citename{Binney \& Tremaine }1987]{binn87}
Binney J., Tremaine S., 1987,
Galactic Dynamics. Princeton Univ. Press, Princeton

\bibitem[\protect\citename{Caputo \& Castellani }1984]{capu84}
Caputo F., Castellani V., 1984,
MNRAS, 207, 185

\bibitem[\protect\citename{Chernoff \& Shapiro }1987]{cherno87}
Chernoff D. F., Shapiro S. L., 1987
ApJ, 322, 113

\bibitem[\protect\citename{Fall \& Rees }1977]{fall77}
Fall S.~M., Rees M.~J., 1977,
MNRAS, 181, 37p

\bibitem[\protect\citename{Fall \& Rees }1985]{fall85}
Fall S.~M., Rees M.~J., 1985,
ApJ, 298, 18

\bibitem[\protect\citename{Frenk et al. }1988]{fren88}
Frenk, C. S., White S. D. M., Davis M., Efstathiou G., 1988,
ApJ, 327, 507

\bibitem[\protect\citename{Funato, Makino \& Ebisuzaki }1993]{funa93}
Funato Y., Makino J., Ebisuzaki T., 1993,
PASJ, 45, 289

\bibitem[\protect\citename{Gnedin \& Ostriker }1997]{gned97}
Gnedin O. Y., Ostriker J. P., 1997,
ApJ, 474, 223

\bibitem[\protect\citename{Makino \& Hut }1997]{makino-j97}
Makino J., Hut P., 1997,
ApJ, 481, 83

\bibitem[\protect\citename{Mihalas \& Binney }1981]{miha81}
Mihalas D., Binney J. J., 1981,
Galactic Astronomy, 2nd edn. Freeman, San Francisco

\bibitem[\protect\citename{Moore, Katz \& Lake }1996]{moor96a}
Moore B., Katz N., Lake G., 1996,
ApJ, 457, 455

\bibitem[\protect\citename{Okazaki \& Tosa }1995]{okaz95}
Okazaki T., Tosa M., 1995,
MNRAS, 274, 48

\bibitem[\protect\citename{Okumura et al. }1993]{okum93}
Okumura S. K. et al., 1993,
PASJ, 45, 329

\bibitem[\protect\citename{Ostriker, Spitzer \& Chevalier }1972]{ostr72}
Ostriker J. P., Spitzer L., Chevalier R. A., 1972,
ApJ, 176, L51

\bibitem[\protect\citename{Peebles }1980]{peeb80}
Peebles P. J. E., 1980,
The Large-Scale Structure of the Universe. 
Princeton Univ. Press, Princeton

\bibitem[\protect\citename{Peebles }1993]{peeb93}
Peebles P. J. E., 1993,
Principles of Physical Cosmology,
Princeton Univ. Press, Princeton

\bibitem[\protect\citename{Richstone \& Malumath }1983]{rich83}
Richstone D.~O., Malumath E.~M., 1983,
ApJ, 268, 30

\bibitem[\protect\citename{Spitzer }1958]{spit58}
Spitzer L., 1958,
ApJ, 127, 17

\bibitem[\protect\citename{Spitzer \& Hart }1971]{spit71}
Spitzer L., Hart M.~H., 1971,
ApJ, 164, 399

\bibitem[\protect\citename{van Albada }1982]{vana82}
{van Albada} T.~S., 1982,
MNRAS, 201, 939

\bibitem[\protect\citename{Vesperini }1997]{vesp97}
Vesperini E., 1997,
MNRAS, 287, 915

\bibitem[\protect\citename{White et al. }1987]{white-s87}
White S. D. M., Davis M., Efstathiou G., Frenk C. S., 1987,
Nat, 330, 451

\end{thebibliography}

\bsp
\label{lastpage}


\begin{table*}
\begin{minipage}{16cm}
\renewcommand{\arraystretch}{1.5}
\begin{center}
\caption{List of parameters and results of our numerical simulations}
\begin{tabular}[p]{|cccccccc|}
\hline
~~$n_k$~~ &~~ $N_c$~~ & ~~~$C$~~~ & ~~~$R_{ini}$~~~&
 virial ratio & ~~~$T_{coll}$~~~ &
~~~$N_{\rm survive}$~~~ & ~~~$N_{\rm bound}$~~~ \\
\hline \hline
 -3 & 2  & 8. & 2.52	& 0.73  &    3.6  &  0/0   &  0/0  \\
 -3 & 4  & 8. & 3.17	& 0.57  &    4.8  &  0/0   &  0/0  \\
 -3 & 8  & 8. & 4.00	& 0.25  &    4.0  &  0/0   &  0/0  \\
 -3 & 16 & 8. & 5.04	& 0.24  &    4.0  &  0/0   &  0/0  \\
 -3 & 32 & 8. & 6.35	& 0.14  &    3.6  &  0/0   &  0/0  \\
 -3 & 64 & 8. & 8.00	& 0.099 &    3.6  &  0/0   &  0/0  \\
 -3 & 128& 8. & 10.1	& 0.063 &    3.4  &  2/2   &  0/0  \\
 -3 & 256& 8. & 12.7	& 0.040 &    3.3  &  0/6   &  0/0  \\
\hline

 -2 & 2  & 11.3 & 2.83	& 0.75  &    4.2  &  0/0   &  0/0  \\
 -2 & 4  & 16.  & 4.00	& 0.62  &    6.6  &  0/0   &  0/0  \\
 -2 & 8  & 22.6 & 5.66	& 0.43  &    6.5  &  0/0   &  0/0  \\
 -2 & 16 & 32.  & 8.00	& 0.32  &    7.9  &  2/0   &  0/0  \\
 -2 & 32 & 45.3 & 11.3	& 0.23  &    8.4  &  0/2   &  0/0  \\
 -2 & 64 & 64.  & 16.0	& 0.18  &   10.0  &  4/2   &  1/0  \\
 -2 & 128& 90.5 & 22.6	& 0.13  &   11.2  & 12/10  &  2/0  \\
 -2 & 256& 128  & 32.0	& 0.095 &   13.2  & 25/30  &  0/4  \\
\hline

 -1 & 2  & 16.   & 3.17	& 0.77  &    4.8  &  0/0   &  0/0  \\
 -1 & 4  & 32.   & 5.04	& 0.67  &    9.2  &  0/0   &  0/0  \\
 -1 & 8  & 64.   & 8.00	& 0.51  &   10.9  &  0/0   &  0/0  \\
 -1 & 16 & 128.  & 12.7	& 0.43  &   15.0  &  0/1   &  0/0  \\
 -1 & 32 & 256.  & 20.2	& 0.35  &   19.5  &  3/2   &  0/0  \\
 -1 & 64 & 512.  & 32.0	& 0.30  &   28.0  & 18/19  &  8/10 \\
 -1 & 128& 1024. & 50.8	& 0.25  &   37.8  & 37/35  & 13/17 \\
 -1 & 256& 2048. & 80.6	& 0.21  &   52.0  & 80/67  & 23/24 \\
\hline

  0 & 2  & 22.627 & 3.56 & 0.79  &    5.6  &  0/0   &  0/0  \\
  0 & 4  & 64     & 6.35 & 0.72  &   12.9  &  0/0   &  0/0  \\
  0 & 8  & 181.01 & 11.3 & 0.60  &   18.1  &  0/0   &  0/0  \\
  0 & 16 & 512    & 20.2 & 0.55  &   30.5  &  1/0   &  1/0  \\
  0 & 32 & 1448.1 & 35.9 & 0.49  &   46.0  &  5/4   &  4/2  \\
  0 & 64 & 4096   & 64.0 & 0.46  &   79.2  & 24/23  & 18/11 \\
  0 & 128& 11585. & 114.0& 0.43  &  126.7  & 70/54  & 48/35 \\
\hline
\end{tabular}
\end{center}
\medskip

Each column means: (1)$n_k$ the index of the power spectrum, (2)$N_c$
the number of clumps, (3) $C$ filling factor, (4) $R_{ini}$ initial
radius of the cluster (5) the initial virial ratio, (6) $T_{coll}$ the
time of the first collapse, (7)$N_{\rm survive}$ the number of clumps which
have a bound core at the final state for RUN1 and RUN2, (8) $N_{\rm
bound}$ the number of surviving clumps which is gravitationally bound in
the merged halo for RUN1 and RUN2.

\end{minipage}
\end{table*}

\clearpage


\end{document}